\documentclass[12pt]{article}
\usepackage[symbol]{footmisc}
\def\correspondingauthor{\footnote{Corresponding author.  }}
\usepackage[left=2.5cm,top=2.5cm,right=2.5cm,bottom=2.5cm]{geometry}
\usepackage{amsmath, amssymb}
\usepackage{graphicx}
\usepackage{graphics}
\usepackage{epstopdf}
\usepackage{subfigure}
\usepackage{amsfonts}
\usepackage{sectsty}
\usepackage{sectsty}
\usepackage{hyperref}
\usepackage{cite}
\usepackage{multicol}

\begin{document}
	\begin{center}
		\large{\bf{Physical Parameters for Stable $f(R)$ Models}} \\
		\vspace{10mm}
		\normalsize{Gauranga C. Samanta$^{1}$ and Nisha Godani$^{2}$\correspondingauthor{}}\\
			\normalsize {gauranga81@gmail.com, nishagodani.dei@gmail.com}
	\\
		\normalsize{$^{1}$Department of Mathematics, BITS Pilani K K Birla Goa Campus, India\\
			$^{2}$Department of Mathematics, Institute of Applied Sciences and Humanities\\ \vspace{-1mm}GLA University, Mathura, Uttar Pradesh, India
 }
	\end{center}

	\begin{abstract}
	 Nojiri \&  Odintsov \cite{noj1} and  Hu \&  Sawicki \cite{hu} have studied non-linear functions in modified gravity that explain the cosmic acceleration without cosmological constant, fulfil the conditions of local gravity \& stability and pass the solar system tests. In this paper,  FRW model, a best fitted and fruitful mathematical model of the physical universe \cite{rob, hub, alph, pen} is studied in the context of these non-linear functions. The cosmological implications such as  Hubble parameter, deceleration parameter, jerk parameter, matter density and the effective equation of state parameter of the universe are plotted with respect to redshift. Subsequently, the age of the universe is predicted in $f(R)$ gravity. All are found to represent the features of present phase of the universe.
		\end{abstract}
	
	\textbf{Keywords:}  $f(R)$ gravity; FRW model; Deceleration parameter; Hubble parameter; Jerk parameter\\

\textbf{PACS Numbers:} 98.80.k, 98.80.Es
	\section{Introduction}
    
    During the last few decades, the observational results of Supernova have declared the expansion of the universe to be accelerating \cite{riess1, perl, ben, riess2}. Consequently, it has motivated towards the theories containing the cosmological constant which were first introduced by Einstein in his theory of cosmological evolution and after some time, he found it to be a great mistake. However, at present, it has been clear that Einstein was not wrong.

   Now, it is not an easy task to find a correct model explaining this late time acceleration. In literature, many efforts have been put in this direction and several models are developed. In the development of these models, two approaches have been taken into account. The first approach makes the use of scalar field which corresponds to a modification of the energy momentum tensor in Einstein equation. The models based on this approach are called quintessence models. On the other hand, the second approach corresponds to the modification of the gravitational theory in comparison of Einstein's general relativity. A simplest theory using the second approach is known as a $f(R)$ theory of gravity  in which the term, $R$, Ricci scalar, appearing in the action is replaced by general function $f(R)$.
       For deriving the field equations of this theory, the two formalisms have been found. In the first one, the metric tensor varies with respect to the metric $g_{\mu\nu}$ and the affine connection $\Gamma_{\mu\nu}^{\gamma}$ which is a function of $g_{\mu\nu}$. However, in the second form which is called Palataini formalism, when the action is varied, $\Gamma_{\mu\nu}^{\gamma}$ and $g_{\mu\nu}$ are considered independent. Starobinsky \cite{star1} proposed a significant  $f(R)$ model with $f(R) = R + \alpha R^2$, where $\alpha>0$, to explain inflationary era of the universe. Subsequently, various other theories such as $f(R,T)$ gravity, $f(G)$ gravity, $f(R, T, Q)$ gravity, Eintein-$\Lambda$ gravity etc. are also introduced in literature and studied in  different aspects \cite{harko, Samantagc, Samantagc1, Samantagc2, Elizalde, Yousaf7, Hussain5, Shabani, Azmat, Samantab, Samantac, Nisha1, Nisha2}. Yousaf et al. \cite{Yousaf1} studied the evolutionary behaviors of compact objects using structure scalars in $f(R,T)$ theory of gravity.
       They considered the spherical geometry coupled with heat and
       radiation emitting shearing viscous matter configurations, constructed structure scalars from the orthogonal decomposition of the Riemann curvature tensor and explored the influence of $f(R,T)$ on dynamics of radiating spherical fluids. Subsequently,
       Yousaf et al. \cite{Yousaf2} studied the distribution of matter
       configuration for a self-gravitating spherical star and examined irregularity factors for dust, isotropic and anisotropic fluids in two regimes in  $f(R, T)$ gravity.
       Bamba et al. \cite{Yousaf3} considered flat FLRW model in the framework of $f(G)$ gravity. They investigated energy conditions and found the viability bounds. Using the recent  values of parameters: Hubble, deceleration, jerk and snap, they obtained the regions satisfying the null and weak energy conditions.
       Yousaf \cite{Yousaf5} considered non-static and non-diagonal cosmic stellar filament in the presence of cosmological constant and studied the effect of expansion-free condition on exact analytical solutions.
       Yousaf et al. \cite{Yousaf4} studied the stability of self gravitating
       celestial body using the background of $f(R, T, Q)$ gravity
       and investigated hydrodynamical equation and
       instability conditions with both N and pN approximations.

    In 1998, the dark energy search puts forward an idea of the gravity modification.  Several models in this regard have been studied in $f(R)$ theory.   Nojiri and  Odintosov \cite{noj1} used the function $f(R)$ including both positive and negative powers of $R$ to obtain both early inflation and late time acceleration. Sotiriou \cite{sotiriou} studied the conditions for $f(R)$ theory of gravity and scalar tensor  theory to be equivalent and explored its implications. Ali et al. \cite{Gannouji} investigated the viable cosmological models in $f(R)$ gravity. Ganguly et al. \cite{Gannouji1} studied the structure of neutron stars in $f(R)$ gravity.  Huang \cite{huang} investigated an $f(R)$ model of inflation. Sharif and Nawazish \cite{sharif} explored warm intermediate inflation in $f(R)$ gravity and obtained the inflation solution in both weak and strong constant regimes and analyze the observational parameters.  Bahamonde et al. \cite{baha} studied the presence of accelerating universe between the frames of $f(R)$ theory of gravity and minimally and non-minimally coupled scalar field theories.
     Felice \cite{felice} reviewed  the applications of $f(R)$ theories and its
     extension to other modified theories of gravity.
    Nojiri et al.\cite{noj2017} also reviewed the recent development in various forms of modified gravity and described the findings on inflations, bounding cosmology and late time acceleration.
 Nojiri and Odintosov \cite{noj} studied $f(R)$, $f(G)$ and $f(R,G)$ models with non-linear gravitational coupling.
Sebastiani  et al. \cite{sebas} investigated class of inflationary scalar potentials in Einstein and Jordan frames. Cognola  et al. \cite{cog2008} studied modified $f(R)$ gravities and described inflation and accelerated expansion. Thakur and Sen \cite{thakur} discussed a non-minimally coupled form of function $f(R)$.  Mukherjee and  Banerjee \cite{ban} studied FRW model with $f(R)$ proportional to $(\lambda + R)^n$ and $exp(\alpha R)$. Guo and Frolov \cite{Guo} explored the cosmological dynamics for a range of $f(R)$  gravity models and studied  the phase-space dynamics, cosmological viability conditions and the cosmological evolution of $f(R)$ gravity models.   Bamba et al. \cite{Bamba1} investigated $f(R)$ gravity models with two forms of scale factor and analyzed occurrence of bounce. Amani \cite{Amani} studied $f(R)$ model in Friedmann–Lemaître–Robertson–Walker (FLRW) framework and explored the nature of bouncing cosmology in $f(R)$ gravity. He also investigated bouncing conditions and obtained late time acceleration. Zubair and Abbas \cite{Zubair} used the solution of Krori and Barua to the anisotropic distribution and studied the formation of compact stars. They computed the constants of Krori and Barua solution and discussed energy conditions, stability, regularity of matter components etc. Yousaf et al. \cite{Yousaf6} studied the effect of Palatini $f(R)$ terms for inhomogeneity factors of spherical relativistic systems. They explored the evolution of  Lema\^{\i}tre-Tolman-Bondi dynamical model with respect to tilted and non-tilted observers for specific types of fluid distribution.
 Subsequently, several authors \cite{Makarenko, Nojiri2014, Bamba2014, Bamba2017, Odintsov2017, Oikonomou2017, Capozziello2018, Odintsov2018,  godani, godani1, Samanta2019} discussed cosmological models in $f(R)$ gravity from different aspects.

At the present time, the universe is passing through the phase of cosmic acceleration which can be well described by general relativity by invoking dark energy. For this, the cosmological constant is found to be a standard and simplest possibility. The smallest estimates for its value are of order 55 \cite{carroll1, peeble}. This led to several other possibilities that considers dark energy associated with a new scalar field \cite{wet, ratra, cald,arm1, arm2, arm3, mersini}.  But these possibilities have also many drawbacks. This puts forward an idea of the gravity modification. Several researchers have made attempts to generalize the action of general relativity and study various $f(R)$ models explaining early inflation or a late time acceleration. 
Since general relativity is based upon Einstein Hilbert action with Lagrangian
density $\sqrt{-g}R$, a natural generalization is the addition of the terms proportional to $\sqrt{-g}R^n$, where $n$ is a constant.
 Caroll et al. \cite{carroll} introduced a gravitational alternative for dark energy by  modifying the Einstein action by adding the term $\dfrac{1}{R}$ which dominates at low curvature. The explanation of present cosmic acceleration through the gravitational foundation is more natural. But it was found to be instable from such theory. This indicated some further modifications.

 Nojiri and  Odintsov \cite{noj1} proposed a new model with Einstein action containing the function both positive and negative powers of curvature as $f(R)=R + R^2 + \dfrac{1}{R}$. The idea was to unify inflation and dark energy. The term $R^2$ would be dominant at large curvature so in the past and would produce a first acceleration i.e. inflation and the term  $\dfrac{1}{R}$ would dominant at low curvature at present and would produce the second acceleration of the universe i.e. dark energy.
After few years, in order to explore the accelerating phase of the universe,  Hu \&  Sawicki \cite{hu} introduced a non-linear function $f(R)$ that satisfy the following properties: (a) it mimics the $\Lambda$CDM model for high redshift, (b) it shows the accelerating stage for low redshift, (c) it exhibits  enough degrees of freedom to incorporate a wide series of low redshift and (d) it includes the $\Lambda$CDM model in a special case. They defined it as $f(R)=R- \mu R_c \frac{\left(\frac{R}{R_c}\right)^{2n}}{\left(\frac{R}{R_c}\right)^{2n}+1}$ where $n, \mu$ and $R_c$ are constant. For $n=1$, this function is same as the non-linear function introduced by Starobinsky \cite{star} for the explanation of accelerating universe without cosmological constant. The second model is also studied in \cite{Odintsov1, Odintsov2}. In these papers, the present value of Hubble parameter is obtained  equal to 0.07GYrs$^-1$ approximately which is same as the value obtained in this work. Also in \cite{Odintsov1,Odintsov2}, the expansion of the universe is obtained from decelerating phase. However, in this work, the evolution of the universe is found from decelerating phase to accelerating phase and at the present time, the universe is shown  to be in an accelerating phase.

In this paper, two forms of function $f(R)$ are taken into account to draw various cosmological implications in the framework of FRW model. These are defined as (i) $f(R)=R + R^2 + \dfrac{1}{R}$ \cite{noj1} and (ii) $f(R)=R- \mu R_c \dfrac{R^2}{R^2 + R_c^2}$ \cite{hu}. As discussed above, the first model produces early inflation and late time acceleration and the second model describes Newton's law at large scale and effective $\Lambda$CDM cosmology. These significant outcomes have motivated  us to consider models (i) and (ii) in the present work. In the context of these models, the cosmological implications such as  Hubble parameter, deceleration parameter, jerk parameter, matter density, the effective equation of state parameter and the age of the universe are plotted in this work.

\section{$f(R)$ Gravity \& Field Equations}
In this section, $f(R)$ gravity and Einstein's field equations for FRW metric are described  briefly. Throughout dot and dash upon a function denote derivative with respect to cosmic time and Ricci scalar respectively.

	In  $f(R)$ gravity, the   action is defined as
	\begin{equation}\label{action}
	S_G=\dfrac{1}{16\pi}\int[f(R) + L_m]\sqrt{-g}d^4x,
	\end{equation}
	
	where $f(R)$ is a function of Ricci scalar $R$, $L_m$ is the matter Lagrangian density, and  $g$ stands for the determinant of the metric $g_{\mu\nu}$.\\
	
	Varying Eq.(\ref{action}) with respect to the metric $g_{\mu\nu}$, the field equations are
	\begin{equation}
	f'(R)_{\mu;\nu}-\square f'(R)g_{\mu\nu}+f'(R)R_{\mu\nu} -\dfrac{1}{2}f(R))g_{\mu\nu}= -\dfrac{8\pi G}{c^4}T_{\mu\nu},
	\end{equation}	
where $T_{\mu\nu}$ is the stress energy tensor of the matter defined as
\begin{equation}
T_{\mu\nu} = (\rho + p)u_\mu u_\nu - pg_{\mu\nu}
\end{equation}	

such that
\begin{equation}
u^\mu\triangledown_\nu u_\mu=0, u^{\mu}u_\mu=1.
\end{equation}

	The flat FRW metric is
	\begin{equation}
	ds^2=dt^2-a^2(t)[dx^2+dy^2+dz^2],
	\end{equation}
	
	where $a(t)$ denotes the scale factor.\\
	
	 The Einstein's field equations are obtained as
	\begin{eqnarray}
	3\dfrac{\dot{a}^2}{a^2}&=&\dfrac{\rho_m}{f'}+\dfrac{1}{f'}\left(\dfrac{1}{2}(-f+Rf')-3\dot{R}f''\dfrac{\dot{a}}{a}\right),\\
	2\dfrac{\ddot{a}}{a}+\dfrac{\dot{a}^2}{a^2}&=&-\dfrac{1}{f'}\left(3\dot{R}f''\dfrac{\dot{a}}{a}+f'''\dot{R}^2+f''\ddot{R}+\dfrac{1}{2}f-R\dfrac{f'}{2}\right),
	\end{eqnarray}
	
		where $\rho_m$ denotes the matter density and the corresponding pressure is taken as zero.\\
		
		Let $\rho_k\equiv\dfrac{1}{2}(-f+Rf')-3\dot{R}f''\dfrac{\dot{a}}{a}$ and $p_k\equiv 3\dot{R}f''\dfrac{\dot{a}}{a}+f'''\dot{R}^2+f''\ddot{R}+\dfrac{1}{2}f-R\dfrac{f'}{2}$. Then $\rho_k$ and $p_k$ represent the curvature density and pressure respectively.

		Equations (6) and (7) reduce to
	\begin{eqnarray}
	3H^2&=&\dfrac{\rho_m+\rho_k}{f'},\\
	2\dot{H}+3H^2&=&-\dfrac{p_k}{f'}.
	\end{eqnarray}
	
		Equations (8) and (9) give
	\begin{equation} \label {con}
	\dfrac{d}{dt}\left(\dfrac{\rho_m + \rho_k}{f'}\right)+3H\left(\dfrac{\rho_m + \rho_k +p_k}{f'}\right)=0.
	\end{equation}
	
		The matter conservation gives
	\begin{equation}
	\dot{\rho_m}+3H\rho_m=0.
	\end{equation}
	
	Then Eq.(\ref{con}) takes the form
	\begin{equation}
	\dfrac{d}{dt}\left(\dfrac{\rho_k}{f'}\right)+3H\left(\dfrac{\rho_k +p_k}{f'}\right)=\dfrac{\rho_m f''\dot{R}}{f'^2}.
	\end{equation}
	
	The scalar  curvature and its derivative with respect to cosmic time are obtained as
	
	\begin{eqnarray}
		R&=&6\left[\dfrac{\dot{a}^2}{a^2}+\dfrac{\ddot{a}}{a}\right]= 6H\left(2H-
		\dfrac{dH}{dz}(1+z)\right)
	\end{eqnarray}

	\begin{equation}
	\dot{R}=6H^3(j-q-2)=6H(1+z)\left[\left\{H\dfrac{d^2H}{dz^2}+\left(\dfrac{dH}{dz}\right)^2\right\}(1+z)-3H\dfrac{dH}{dz}\right],
	\end{equation}
	where $q$ and $j$ are deceleration and jerk parameters respectively.


		\section{Physical Parameters for $f(R)$ Models}
			The exploration of the dynamics of the universe has become an important part of research for cosmologists. In this regard, Mukherjee and  Banerjee \cite{ban} taken into account two $f(R)$ models proportional to $(\lambda + R)^n$ and $exp(\alpha R)$ and plotted deceleration and equation of state parameters with respect to redshift for different values of model parameters.  They obtained present accelerating phase of the universe  and found the value of the equation of parameter to  lie between -0.8 t0 -1. In this work, two other significant $f(R)$ models are considered  in the framework of FRW model. The first one was defined by Nojiri \& Odintsov \cite{noj1} as  $f(R)=R + R^2 + \dfrac{1}{R}$ \cite{noj1} and the second one was proposed by Hu \& Sawicki \cite{hu} as $f(R)=R- \mu R_c \dfrac{R^2}{R^2 + R_c^2}$ \cite{hu}. In this section, the physical parameters which include Hubble parameter, deceleration parameter, jerk parameter, matter density, effective equation of state parameter and age of the universe are plotted with respect to redshift for these two $f(R)$ gravity models. To find the numerical plots of  these physical parameters, the present values of deceleration parameter and jerk parameter are taken to be equal to -0.81  and 2.16 respectively \cite{repeti}.\\

		From Equations (8), (13) \& (14),
	
		 \begin{equation}\label{diff}
		 \dfrac{d^2H}{dz^2}=-\dfrac{1}{H}\left(\dfrac{dH}{dz}\right)^2+\dfrac{3}{(1+z)}\dfrac{dH}{dz}-\dfrac{3f'\left[(1+z)\dfrac{dH}{dz}-H^2\right]+\left(\dfrac{f}{2}-\rho_{m0}\right)}{18f''H^3(1+z)^2}.
		 \end{equation}
		
		
		  Eq. (\ref{diff}) is a non-linear differential equation for Hubble parameter $H$ as a function of redshift $z$.  Hubble parameter is scaled as $\dfrac{H}{H_0}$ so that its value at present is unity.

	 	
%
		
		 	The deceleration parameter  is
		 \begin{eqnarray}
		 q&=&\dfrac{a\ddot{a}}{\dot{a}^2}\nonumber\\
		 &=&-1+(1+z)\dfrac{1}{H}\dfrac{dH}{dz}.\end{eqnarray}
		
		Jerk parameter is
		\begin{eqnarray}
		j&=&\dfrac{\dddot{a}}{aH^3}\nonumber \\
		&=& 1+\dfrac{3\dot{H}}{H^2}+\dfrac{\dddot{H}}{H^3}.
		\end{eqnarray}
		
		 The effective equation of state parameter is given by
		 		 \begin{eqnarray}
		 w_{eff}&=&\dfrac{p_k}{\rho_k + \rho_m}\nonumber \\
		 &=& -1+\dfrac{2}{3}(1+z)\dfrac{1}{H}\dfrac{dH}{dz}.
		 \end{eqnarray}
		
		 The age of the universe is given by
		 \begin{eqnarray}\label{t}
		 t_0&=&\int_0^{t_0}dt\nonumber\\
		 &=&\int_{0}^{\infty}\dfrac{dz}{H(z)(1+z)}.
		 	 \end{eqnarray}

\textbf{Case-I: $f(R)=R+R^2+\dfrac{1}{R}$}

In this case, the function $f(R)$ is taken as  $f(R)=R+R^2+\dfrac{1}{R}$ which is free from parameters. The term $\dfrac{1}{R}$ dominates at low curvature and produces the cosmic acceleration. This model is stable and satisfy local gravity constraints \cite{noj1}.

		 \begin{figure}[h!]
		 	\begin{center}
		 		\includegraphics[height=8.4cm,width=10cm]{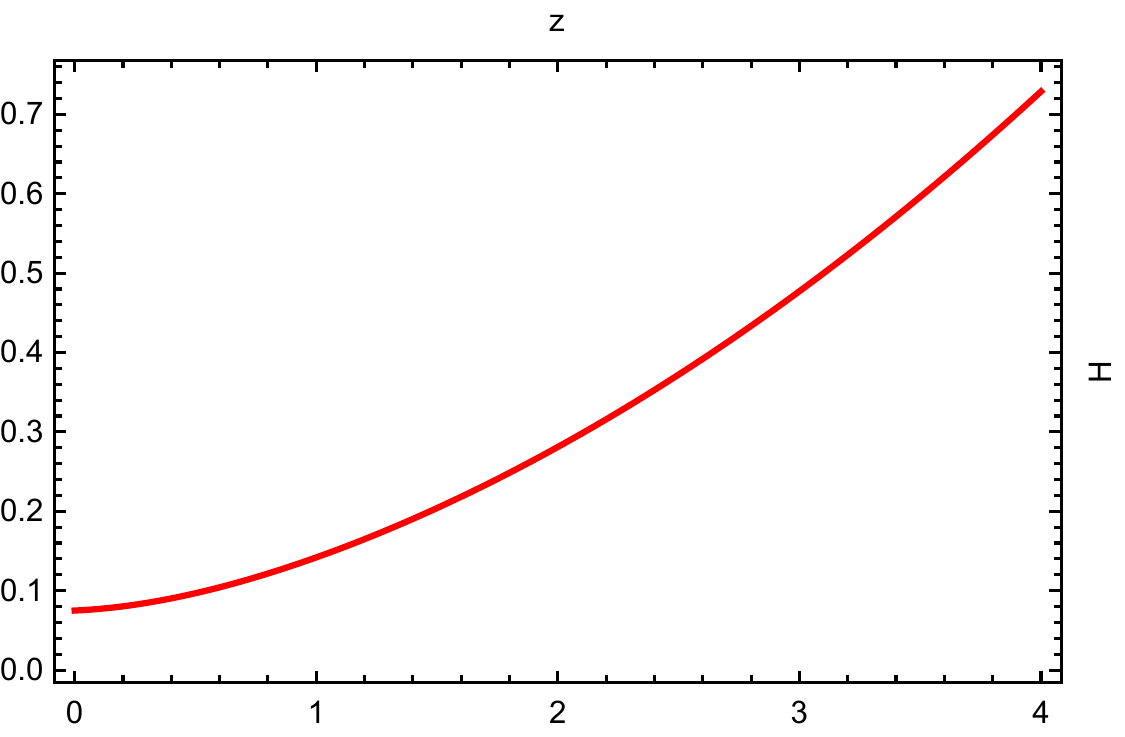}
		 		\caption{Hubble parameter $H$ verses redshift $z$}
		 	\end{center}
		 \end{figure}

		 \begin{figure}[h!]\label{qz}
		 	\begin{center}
		 		\includegraphics[height=8.4cm,width=10cm]{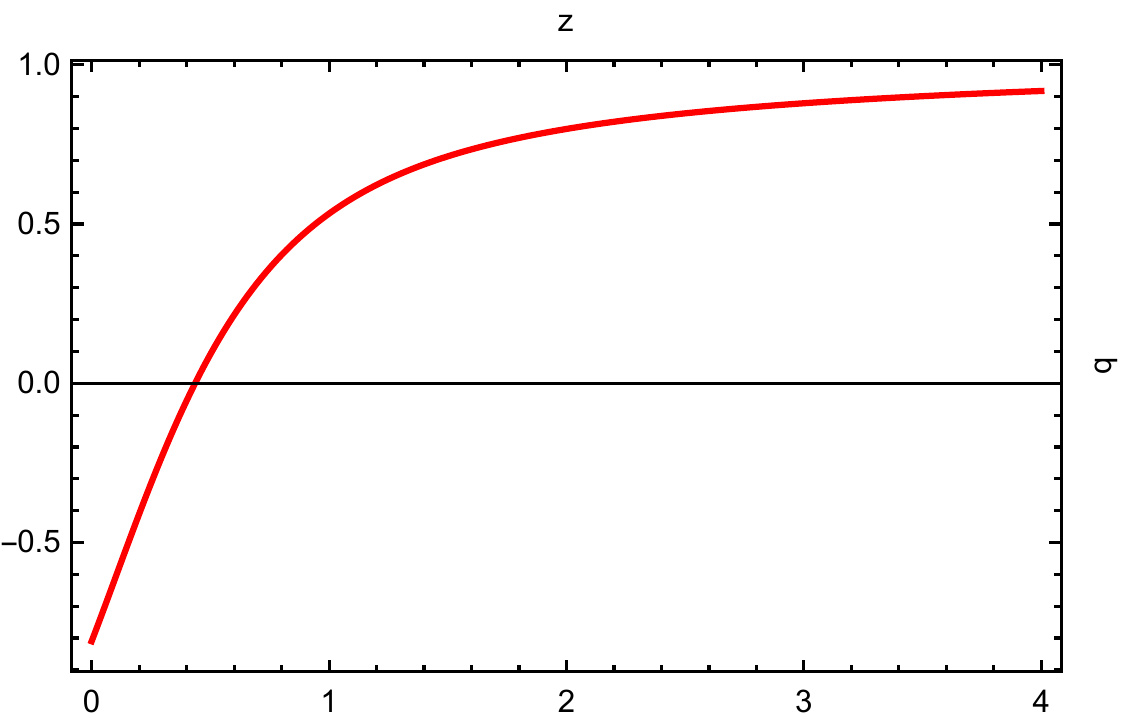}
		 		\caption{The  deceleration parameter $q$ versus redshift $z$}	\end{center}
		 \end{figure}
	
	\begin{figure}[h!]\label{}
		\begin{center}
			\includegraphics[height=8.4cm,width=10cm]{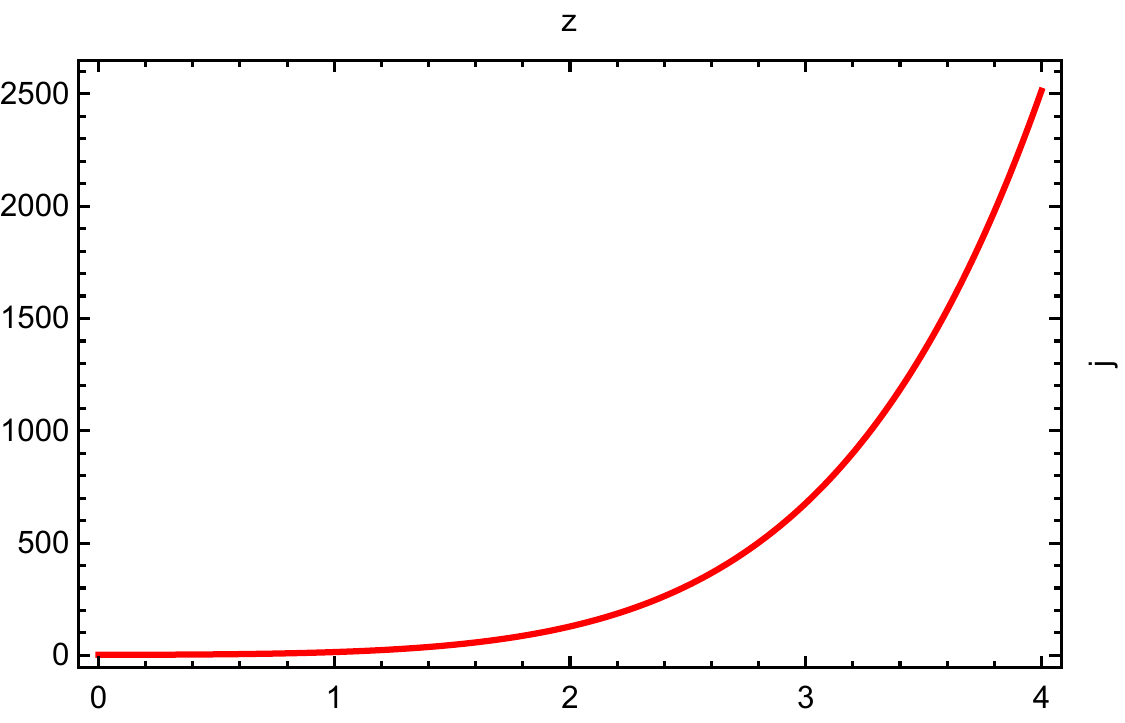}
			\caption{The  jerk parameter $j$ versus redshift $z$}	\end{center}
	\end{figure}
		
		 	\begin{figure}[h!]\label{}
		 	\begin{center}
		 		\includegraphics[height=8.4cm,width=10cm]{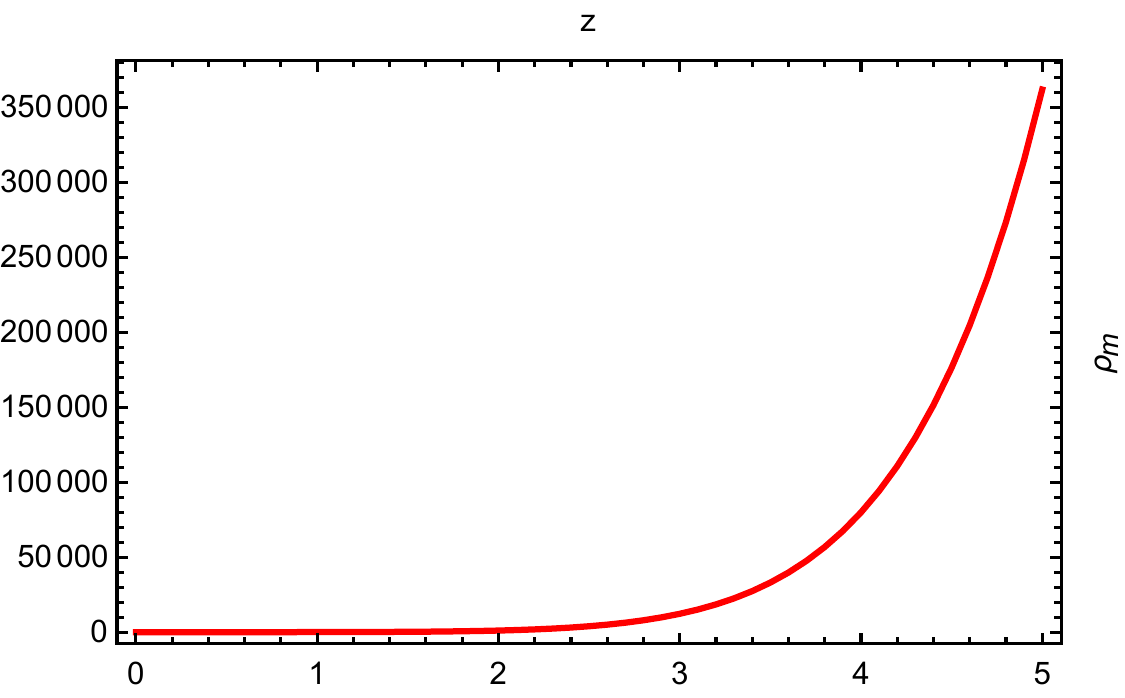}
		 		\caption{Matter density $\rho_m$ versus redshift $z$}	\end{center}
		 \end{figure}
	
		 \begin{figure}[h!]
		 		 	\begin{center}
		 		\includegraphics[height=8.4cm,width=10cm]{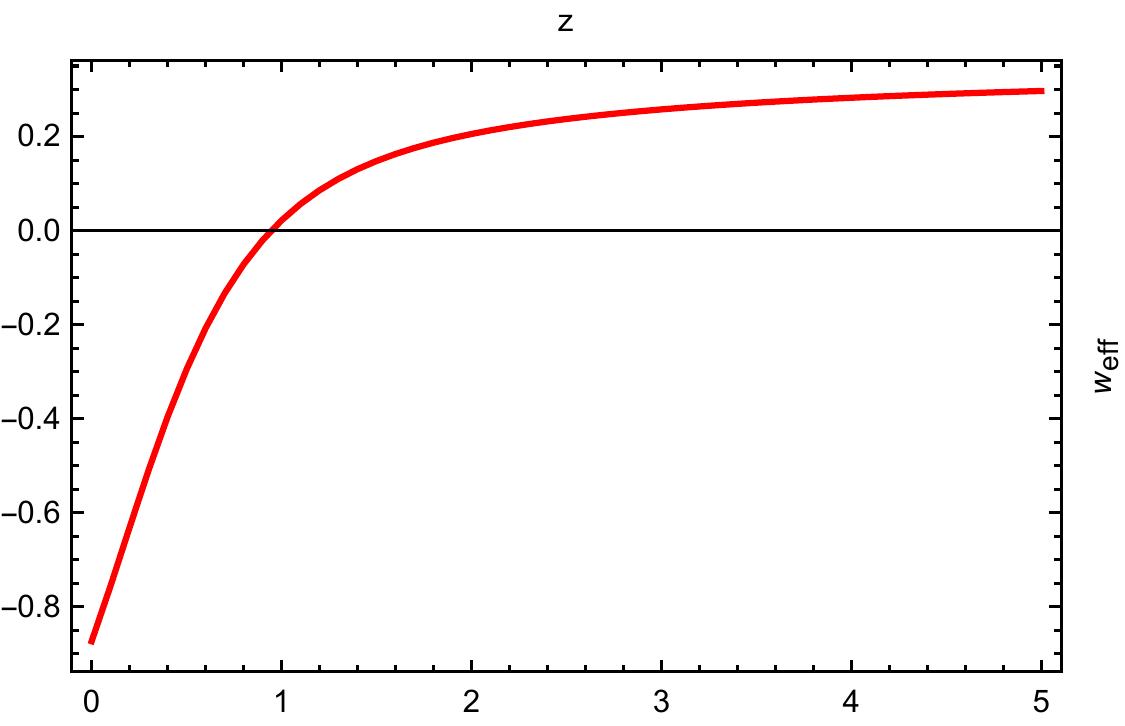}
		 		\caption{The equation of state parameter $w_{eff}$ versus redshift $z$}
		 	\end{center}
		 \end{figure}
	
	 		
	  \begin{figure}[h!]
	 	\begin{center}
	 		\includegraphics[height=8.4cm,width=10cm]{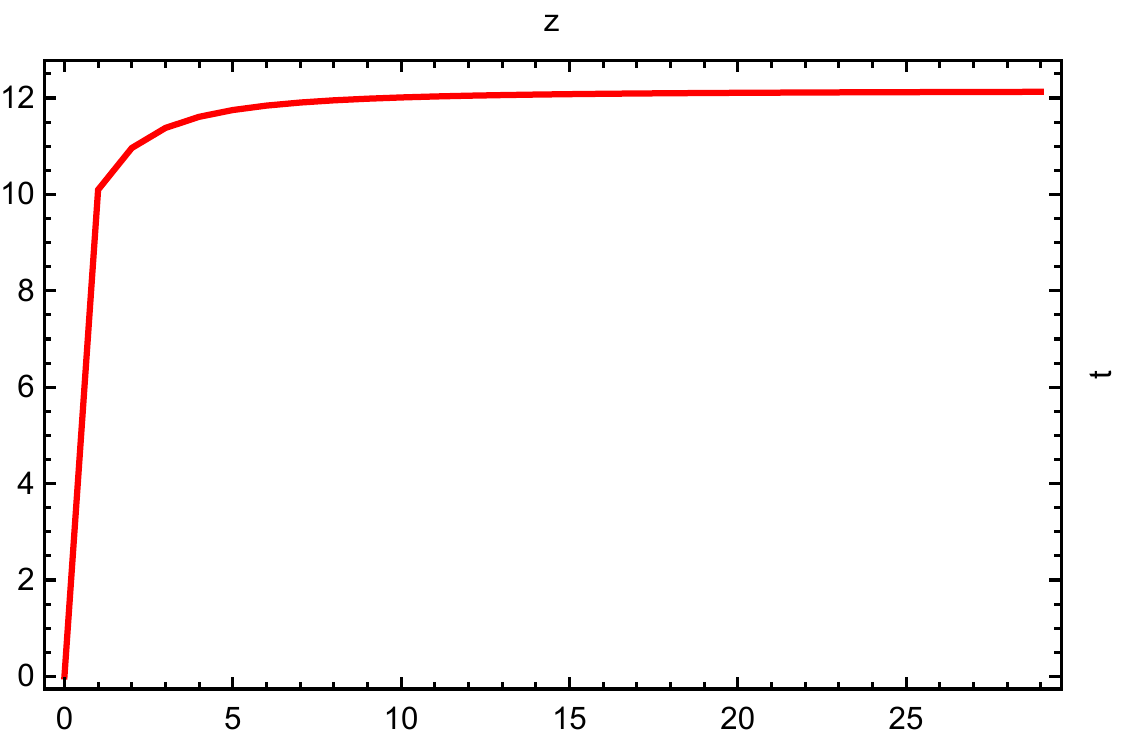}
	 		\caption{Age $t$ versus redshift $z$}	
	 	\end{center}
	 \end{figure}
\clearpage
	 \textbf{		Case-II: $f(R)=R- \mu R_c \dfrac{R^2}{R^2 + R_c^2}$}
	
	 In this case, the function $f(R)$ is taken as  $f(R)=R- \mu R_c \dfrac{R^2}{R^2 + R_c^2}$ where $\mu$ and $R_c$ are constants. This was first introduced by W. Hu \& I. Sawick \cite{hu} in the form $f(R)=R- \mu R_c \dfrac{\left(\dfrac{R}{R_c}\right)^{2n}}{\left(\dfrac{R}{R_c}\right)^{2n}+1}$.
	 For $n=1$, $\mu\geq 8\sqrt{3}/9$ and $R_c\leq 5.7735 \times 10^{-30}$. Here,  $\mu=1.6$ and $R_c = 5.7\times 10^{-30}$ is taken.
	   This model is stable and satisfy local gravity constraints.  It approaches to $\Lambda CDM$ model for $R\gg R_c$.
	
	  \vspace{2cm}
	
	 \begin{figure}[h!]
		\begin{center}
			\includegraphics[height=8.4cm,width=10cm]{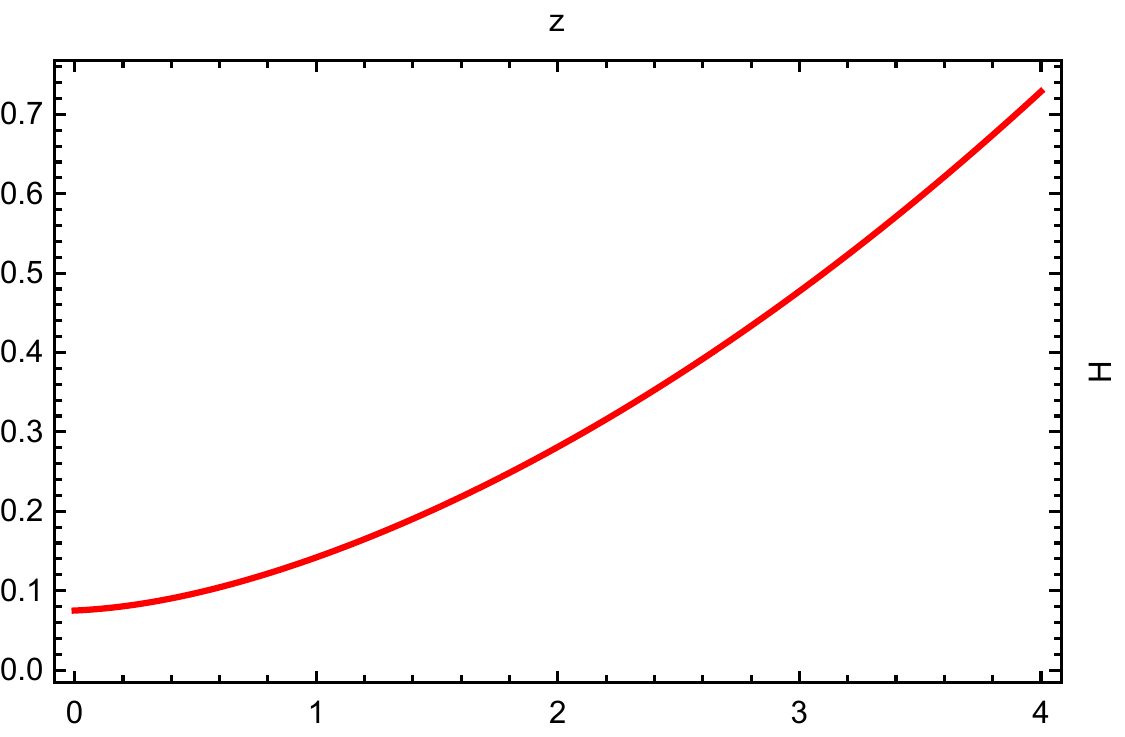}
			\caption{Hubble parameter $H$ verses redshift $z$}
		\end{center}
	\end{figure}

	\begin{figure}[h!]\label{qz}
		\begin{center}
			\includegraphics[height=8.4cm,width=10cm]{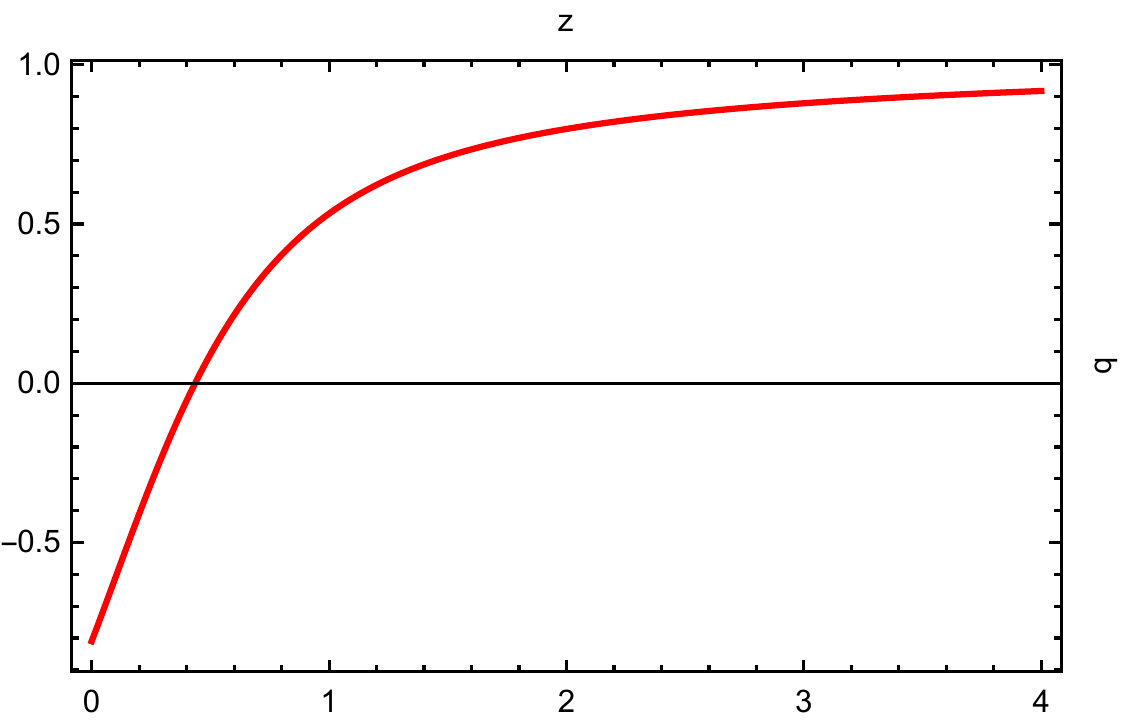}
			\caption{The  deceleration parameter $q$ versus redshift $z$}	\end{center}
	\end{figure}
	
	\begin{figure}[h!]\label{}
		\begin{center}
			\includegraphics[height=8.4cm,width=10cm]{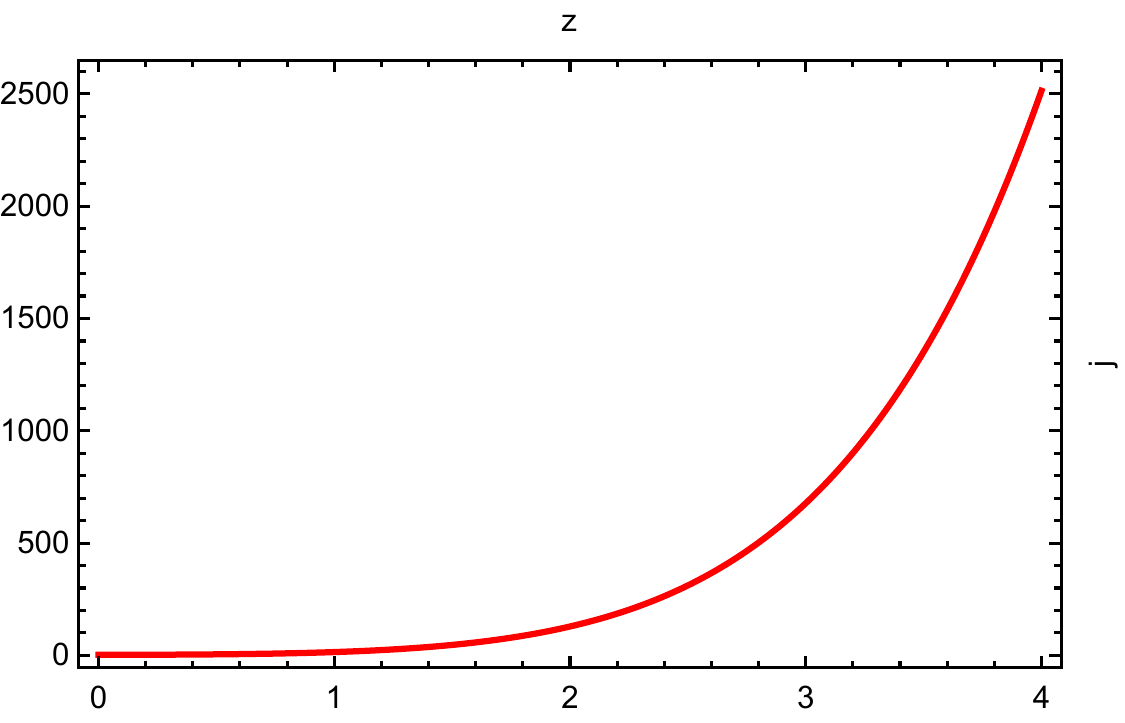}
			\caption{The  jerk parameter $j$ versus redshift $z$}	\end{center}
	\end{figure}
	
			\vspace{2cm}
			
	\begin{figure}[h!]\label{}
		\begin{center}
			\includegraphics[height=8.4cm,width=10cm]{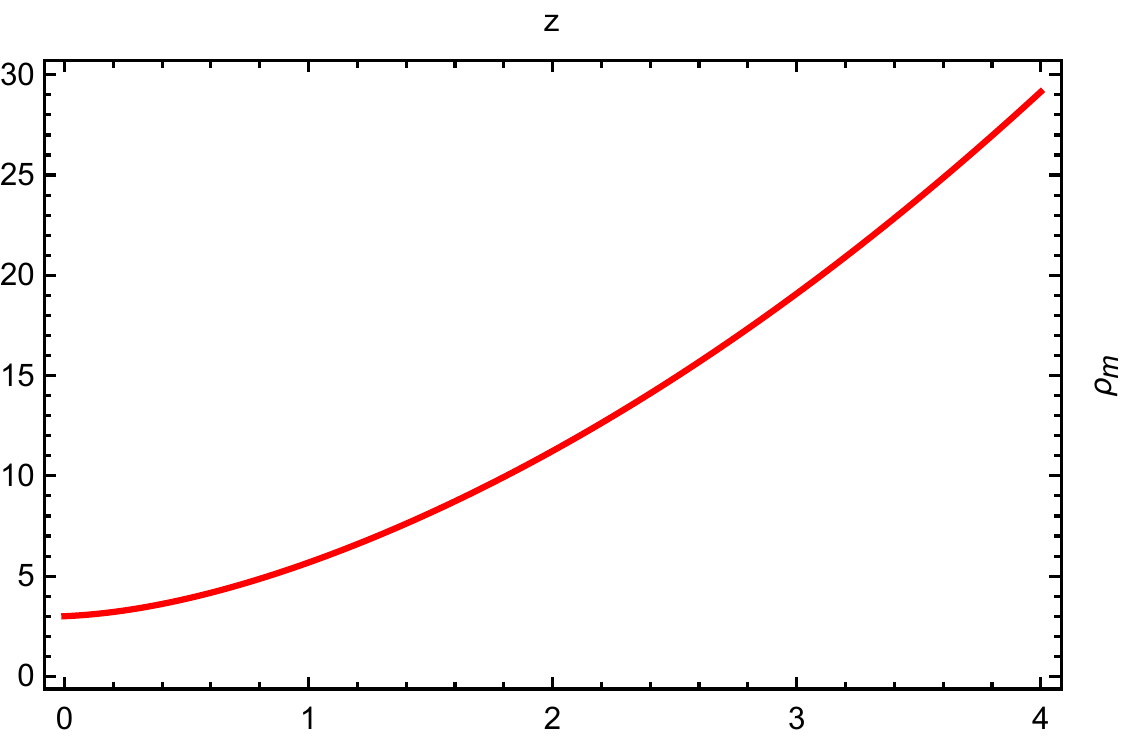}
			\caption{Matter density $\rho_m$ versus redshift $z$}	\end{center}
		\end{figure}
			\vspace{2cm}
	\begin{figure}[h!]
		\begin{center}
			\includegraphics[height=8.4cm,width=10cm]{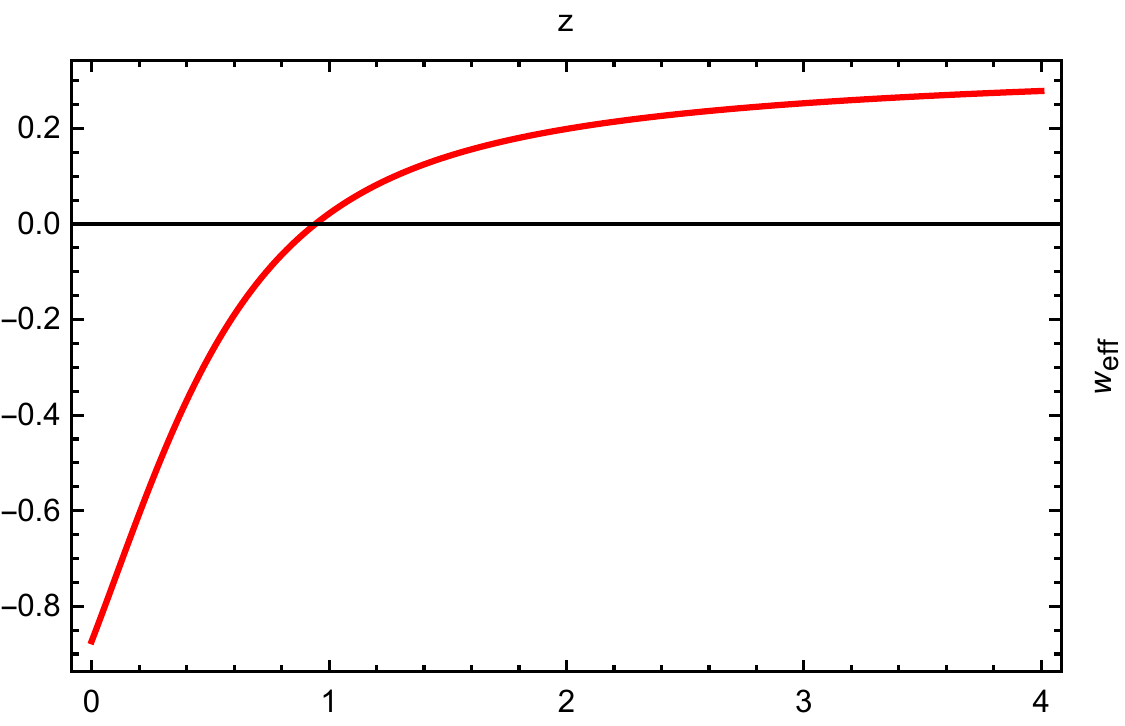}
			\caption{The equation of state parameter $w_{eff}$ versus redshift $z$}
		\end{center}
	\end{figure}
		\vspace{2cm}
			
	\begin{figure}[h!]
		\begin{center}
			\includegraphics[height=8.4cm,width=10cm]{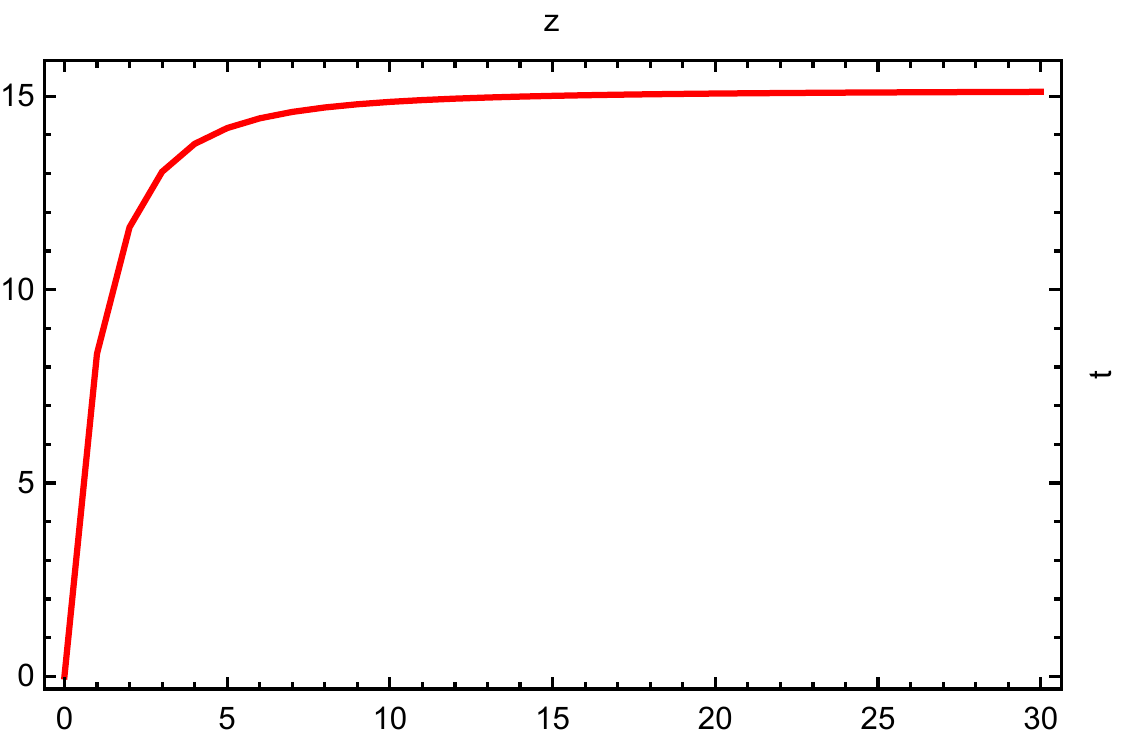}
			\caption{Age $t$ versus redshift $z$}	
		\end{center}
	\end{figure}


		\vspace{2cm}

\section{Results \& Discussion}
Various parameters describing the evolution of the universe are plotted with respect to redshift in the context of two $f(R)$ gravity models in the previous section (Figs. (1-12)). These include Hubble parameter, deceleration parameter, jerk parameter, matter density, the effective equation of state parameter and age of the universe. The results obtained are discussed below:

In Figures (1) and (7), Hubble parameter is plotted with respect to redshift $z$.  In both cases, it tends to infinity as $z\rightarrow \infty$ and its present value is found to be equal to  0.07 GYrs$^{-1}$ which is consistent with the current observational data \cite{wmap}. In Figures (2) and (8), deceleration parameter is plotted with respect to redshift $z$ for both cases. It depicts a change of the universe from a deceleration stage to an acceleration stage. At $z=0$, it is nearly equal to -0.8, in each case, which is also very much closed to recent experimental value \cite{repeti}. In Figures (3) and (9), jerk parameter is drawn with respect to redshift. In the first case, its present values are obtained as 2.16 and in the second case, it is found as 2.43 which is very much closed to the current measured value \cite{repeti}.
In Figures (4) and (10), the matter density $\rho_m$ is plotted with respect to redshift for both cases. It goes to infinity as $z$ goes to infinity.
 In Figures (5) and (11),  the effective equation of state parameter $w$ is plotted with respect to redshift $z$. In case-I, its value is obtained to lie between -0.9 and 0.3 and in case-II, it lies between -0.3 and 0.3 which confirms the latest observational estimates. In Figures (6) and (12), the cosmic time $t$ is plotted with respect to redshift. In the first case, it is observed that $t$ tends to 12.13 as $z$ tends to infinity, while in the second case, it is found to approach 15.11. Thus, the age of the universe is obtained as 12.13 GYrs and 15.11 GYrs in case-I and-II respectively which is closed to the present age of the universe according to WMAP data \cite{wmap}. Hence, all the physical parameters drawn are observed to be consistent with the current observational results.

\section{Conclusion}
The present paper is devoted to the study of  FRW model in $f(R)$ gravity with two stable  $f(R)$ functions proposed in \cite{noj1,hu} explaining the cosmic acceleration and satisfying the solar system experiments. For both functions,
Hubble parameter, deceleration parameter, jerk parameter, matter density, the effective equation of state parameter and age of the universe are plotted in terms of redshift. Each cosmological implication is found to possess the value compatible with the observational results. The evolution of the universe is observed from decelerating stage to accelerating stage. Thus, the models considered are significant for the exploration of the evolution of our universe.

\section*{Acknowledgement}
The authors are very much thankful to the reviewers and editors for their constructive comment to improve the work significantly. The authors are also thankful to Prof. G.K. Goswami, Kalyan P. G. College,
Bhilai, India, for his motivation to carry out this research work.


\begin{thebibliography}{000}
	\bibitem{noj1} S Nojiri and S D Odintosov \textit{Physical Review D} \textbf{68} 123512 (2003).
	\bibitem{hu} W Hu and I Sawicki \textit{Physical Review D} \textbf{76} 064004 (2007).
	\bibitem{rob} H P Robertson \textit{Astrophysical Journal} \textbf{82} 248 (1935).
	\bibitem{hub}E Hubble \textit{Proceedings of the National Academy of Sciences} \textbf{15} 168 (1929).
	\bibitem{alph}R A Alpher, H A Bethe and G Gamow \textit{Physical Review D} \textbf{73} 80 (1948).
	\bibitem{pen} A A Penzias and R W Wilson \textit{Astrophysical Journal} \textbf{142} 419 (1965).
	\bibitem {riess1}A G Riess et al. \textit{Astrophysical Journal} \textbf{116} 1009 (1998).
	\bibitem {perl}S Perlmutter et al. \textit{Astrophysical Journal} \textbf{517} 565 (1999).
	\bibitem {ben}C L Bennet et al. \textit{The Astrophysical Journal Supplement Series} \textbf{148} 1 (2003).
	\bibitem {riess2}A G Riess et al. \textit{Astrophysical Journal} \textbf{607} 665 (2004).
		
\bibitem{star1}	A A Starobinsky \textit{Physics Letters B} \textbf{91} 99 (1980).


\bibitem{harko} T Harko et al  \textit{Physical Review D} \textbf{84} 024020 (2011).

\bibitem{Samantagc}G C Samanta and S N Dhal \textit{Int. J. Theor. Phys.} \textbf{52} 1334 (2013).


\bibitem{Samantagc1}G C Samanta \textit{Int. J. Theor. Phys.} \textbf{52} 2303 (2013).


\bibitem{Samantagc2}G C Samanta \textit{Int. J. Theor. Phys.} \textbf{52}  2647 (2013).





\bibitem{Elizalde}E Elizalde and S I Vacaru \textit{Gen. Relativ. Gravit.} \textbf{47} 64 (2015).
\bibitem{Yousaf7}Z Yousaf, M Z Bhatti and U Farwa \textit{Mon. Not. R. Astron. Soc.} \textbf{464} 4509 (2017).
\bibitem{Hussain5}T Hussain, M Khurshudyan, S Ahmed and A Z Khurshudyan \textit{Int. J. Mod. Phys. D} \textbf{26} 1750155 (2017).

\bibitem{Shabani}  	H Shabani and A H Ziaie \textit{Eur. Phys. J. C} \textbf{78}  397 (2018).




\bibitem{Azmat}  	
H Azmat, M Zubair and I Noureen \textit{Int. J. Mod. Phys. D} \textbf{27}  1750181 (2017).

\bibitem{Samantab}G C Samanta and R Myrzakulov \textit{Chinese J. Phys.} \textbf{55} 1044 (2017).

\bibitem{Samantac} G C Samanta, R Myrzakulov and P Shah \textit{Z. Naturforsch. A} \textbf{72}  365 (2017).
\bibitem{Nisha1}N Godani \textit{Int. J. Geom. Methods Mod. Phys.} \textbf{16} 1950024 (2019).
\bibitem{Nisha2} N Godani \textit{Indian J Phys} (2019). https://doi.org/10.1007/s12648-018-01363-w
\bibitem{Yousaf1} Z Yousaf, K Bamba and M Z Bhatti, \textit{Phys. Rev. D} \textbf{93}  064059 (2016).
\bibitem{Yousaf2}Z Yousaf, K Bamba and M Z Bhatti \textit{Phys. Rev. D} \textbf{93} 064059 (2016).
\bibitem{Yousaf3}K Bamba, M Ilyas, M Z Bhatti and Z Yousaf \textit{Gen. Relativ. Gravit.} \textbf{49} 112 (2017).

\bibitem{Yousaf5} Z Yousaf \textit{Eur. Phys. J. Plus} \textbf{132} 276 (2017).

\bibitem{Yousaf4}  Z Yousaf, K Bamba, M Z Bhatti and U Farwa \textit{Eur. Phys. J. A}  \textbf{54} 122 (2018).



\bibitem{sotiriou} T P Sotiriou \textit{Classical Quantum Gravity} \textbf{23}  5117 (2006).
\bibitem{Gannouji}A Ali, R Gannouji, M Sami and A A Sen \textit{Physical Review D} \textbf{81} 104029 (2010).
\bibitem{Gannouji1}A Ganguly, R Gannouji, R Goswami, and S Ray
\textit{Physical Review D} \textbf{89} 064019 (2014).
\bibitem{huang}Q G Huang \textit{Journal of Cosmology and Astroparticle Physics} \textbf{35} 1402 (2014).
\bibitem{sharif} M Sharif and I Nawazish \textit{Astrophysics and Space Science} \textbf{362} 30 (2017).
\bibitem{baha} S Bahamonde et al. \textit{Physics Letters B} \textbf{766}  225 (2017).
\bibitem{felice}  A D Felice \textit{Living Reviews in Relativity} \textbf{13} 1 (2010).
	\bibitem{noj2017} S Nojiri et al. \textit{Physics Reports} \textbf{692} 1 (2017).
		\bibitem{noj}S Nojiri and S D Odintsov \textit{International Journal of Geometric Methods in Modern Physics} \textbf{4} 115 (2007).
	
	\bibitem{sebas} L Sebastiani et al. \textit{Physical Review D} \textbf{89} 023518 (2013).
	\bibitem{cog2008} G Cognola et al. \textit{Physical Review D} \textbf{77} 046009 (2008).


	
	\bibitem{thakur}S Thakur and A A Sen \textit{Physical Review D} \textbf{88} 044043 (2013).
	\bibitem{ban}A Mukherjee and N Banerjee \textit{Astrophysics and Space Science} \textbf{352} 893 (2014).
\bibitem{Guo}J Guo and A V Frolov \textit{Phys. Rev. D} \textbf{88} 124036 (2013).
\bibitem{Bamba1}K Bamba, A N Makarenko, A N Myagky, S Nojiri and S D Odintsov \textit{JCAP} \textbf{01} 008 (2014).
	\bibitem{Amani} A R Amani \textit{Int. J. Mod. Phys. D} \textbf{25} 1650071 (2016).
\bibitem{Zubair} M Zubair and G Abbas \textit{arXiv:1412.2120v3[physics.gen-ph]} (2016).
\bibitem{Yousaf6}Z Yousaf, K Bamba and M Z Bhatti \textit{Phys. Rev. D} \textbf{95} 024024 (2017).



\bibitem{Makarenko} 	
K Bamba, A N Makarenko, A N Myagky, S Nojiri and S D Odintsov \textit{Journal of Cosmology and Astroparticle Physics} \textbf{1401} 008 (2014).


\bibitem{Nojiri2014} 	
K Bamba, S Nojiri, S D Odintsov and D Saez-Gomez \textit{Physics Letters B} \textbf{730} 136 (2014).


\bibitem{Bamba2014} 	
K Bamba, S Nojiri, S D Odintsov and D Saez-Gomez \textit{Physical Review D} \textbf{90} 124061 (2014).


\bibitem{Bamba2017}  	
Z Yousaf, K Bamba and M Zaeem-ul-Haq Bhatti \textit{Physical Review D} \textbf{95} 024024 (2017).


\bibitem{Odintsov2017} 	
S D Odintsov and V  K Oikonomou \textit{Physical Review D} \textbf{96} 104049 (2017).


\bibitem{Oikonomou2017} 	
S  D Odintsov and V  K Oikonomou \textit{Annals of Physics} \textbf{388} 267 (2018).


\bibitem{Capozziello2018}	
S Capozziello, S Nojiri and S D Odintsov \textit{Physics Letters B} \textbf{781} 99 (2018).

\bibitem{Odintsov2018}  	
S  D Odintsov and V  K Oikonomou \textit{Physical Review D} \textbf{98} 024013 (2018).
\bibitem{godani} N Godani and G C Samanta \textit{Int. J. Mod. Phys. D} \textbf{28} 1950039 (2018).
\bibitem{godani1} G C Samanta, N Godani and K Bamba \textit{arXiv:1811.06834v1[gr-qc]} (2018).


\bibitem{Samanta2019} 	
A V Astashenok, K Mosani, S D Odintsov and G C Samanta  \textit{arXiv:1812.10441[gr-qc]} (2019)


	\bibitem{carroll1} S M Carroll \textit{Living Reviews in Relativity} \textbf{4} 1 (2001).
	\bibitem{peeble}P J Peebles and B Ratra \textit{Reviews of Modern Physics} \textbf{75} 599	(2003).
\bibitem{wet}C Wetterich \textit{Nuclear Physics B} \textbf{302} 668 (1988).
\bibitem{ratra} B Ratra and P J Peebles \textit{Physical Review D} \textbf{37} 3406 (1988).

\bibitem{cald} R R Caldwell, R Dave and P J Steinhardt  \textit{Physical Review Letters} \textbf{80} 1582 (1998).
\bibitem{arm1} C Armendariz-Picon, T Damour and V Mukhanov \textit{Physics Letters B} \textbf{458} 209 (1999).
\bibitem{arm2} C Armendariz-Picon, V Mukhanov and P J Steinhardt
\textit{Physical Review Letters} \textbf{85} 4438 (2000).
\bibitem{arm3} C Armendariz-Picon, V Mukhanov and
P J Steinhardt \textit{Physical Review D} \textbf{63} 103510 (2001).
\bibitem{mersini} L Mersini, M Bastero-Gil and P Kanti \textit{Physical Review D} \textbf{64} 043508 (2001).
\bibitem{carroll}S M Carroll et al. \textit{Physical Review D} \textbf{70} 043528 (2004).
\bibitem{star}A A Starobinsky \textit{JETP Letters} \textbf{86} 157 (2007).
\bibitem{Odintsov1} S Nojiri and  S D Odintsov  \textit{Phys. Lett. B} \textbf{657} 238 (2007).
\bibitem{Odintsov2} S Nojiri and  S D Odintsov \textit{Phys. Rev. D} \textbf{77} 026007 (2008).

\bibitem{repeti} D Rapetti  et al. \textit{Monthly Notices of the Royal Astronomical Society} \textbf{375} 1510 (2007).
\bibitem{wmap} G Hinshaw et al. \textit{Astrophysical Journal Supplement Series} \textbf{208} 19 (2013).
	
\end{thebibliography}
\end{document}